\documentclass[11pt]{article}
\usepackage{amsfonts,amssymb,a4,bm}
\usepackage[all]{xy}
\usepackage{color}

\def\theequation{\thesection.\arabic{equation}}

\newcommand{\qed}{\hfill\rule{3mm}{3mm}}
\newtheorem{teorema}{Theorem}
\newtheorem{lema}{Lemma}

\makeatletter \@addtoreset{equation}{section} \makeatother
\begin{document}


\voffset=-1.5truecm\hsize=16.5truecm    \vsize=24.truecm
\baselineskip=14pt plus0.1pt minus0.1pt \parindent=12pt
\lineskip=4pt\lineskiplimit=0.1pt      \parskip=0.1pt plus1pt

\def\ds{\displaystyle}\def\st{\scriptstyle}\def\sst{\scriptscriptstyle}


\let\a=\alpha \let\b=\beta \let\ch=\chi \let\d=\delta \let\e=\varepsilon
\let\f=\varphi \let\g=\gamma \let\h=\eta    \let\k=\kappa \let\l=\lambda
\let\m=\mu \let\n=\nu \let\o=\omega    \let\p=\pi \let\ph=\varphi
\let\r=\rho \let\s=\sigma \let\t=\tau \let\th=\vartheta
\let\y=\upsilon \let\x=\xi \let\z=\zeta
\let\D=\Delta \let\F=\Phi \let\G=\Gamma \let\L=\Lmbda \let\Th=\Theta
\let\O=\Omega \let\P=\Pi \let\Ps=\Psi \let\Si=\Sigma \let\X=\Xi
\let\Y=\Upsilon\let\L\Lambda



\global\newcount\numsec\global\newcount\numfor
\gdef\profonditastruttura{\dp\strutbox}
\def\senondefinito#1{\expandafter\ifx\csname#1\endcsname\relax}
\def\SIA #1,#2,#3 {\senondefinito{#1#2}
\expandafter\xdef\csname #1#2\endcsname{#3} \else \write16{???? il
simbolo #2 e' gia' stato definito !!!!} \fi}
\def\etichetta(#1){(\veroparagrafo.\veraformula)
\SIA e,#1,(\veroparagrafo.\veraformula)
 \global\advance\numfor by 1
 \write16{ EQ \equ(#1) ha simbolo #1 }}
\def\etichettaa(#1){(A\veroparagrafo.\veraformula)
 \SIA e,#1,(A\veroparagrafo.\veraformula)
 \global\advance\numfor by 1\write16{ EQ \equ(#1) ha simbolo #1 }}
\def\BOZZA{\def\alato(##1){
 {\vtop to \profonditastruttura{\baselineskip
 \profonditastruttura\vss
 \rlap{\kern-\hsize\kern-1.2truecm{$\scriptstyle##1$}}}}}}
\def\alato(#1){}
\def\veroparagrafo{\number\numsec}\def\veraformula{\number\numfor}
\def\Eq(#1){\eqno{\etichetta(#1)\alato(#1)}}
\def\eq(#1){\etichetta(#1)\alato(#1)}
\def\Eqa(#1){\eqno{\etichettaa(#1)\alato(#1)}}
\def\eqa(#1){\etichettaa(#1)\alato(#1)}
\def\equ(#1){\senondefinito{e#1}$\clubsuit$#1\else\csname e#1\endcsname\fi}
\let\EQ=\Eq

\def\V{V}

\def\dpr{\partial}


\def\\{\noindent}
\let\io=\infty

\def\VU{{\mathbb{V}}}
\def\ED{{\mathbb{E}}}
\def\GI{{\mathbb{G}}}
\def\Tt{{\mathbb{T}}}
\def\C{\mathbb{C}}
\def\LL{{\cal L}}
\def\RR{{\cal R}}
\def\SS{{\cal S}}
\def\NN{{\cal M}}
\def\MM{{\cal M}}
\def\HH{{\cal H}}
\def\GG{{\cal G}}
\def\PP{{\cal P}}
\def\AA{{\cal A}}
\def\BB{{\cal B}}
\def\FF{{\cal F}}
\def\TT{{\cal T}}
\def\v{\vskip.1cm}
\def\vv{\vskip.2cm}
\def\gt{{\tilde\g}}
\def\E{{\mathcal E} }
\def\I{{\rm I}}
\def\0{\emptyset}
\def\xx{{\V x}} \def\yy{{\bf y}} \def\kk{{\bf k}} \def\zz{{\bf z}}
\def\ba{\begin{array}}
\def\ea{\end{array}}  \def \eea {\end {eqnarray}}\def \bea {\begin {eqnarray}}
\def\xto#1{\xrightarrow{#1}}

\def\tende#1{\vtop{\ialign{##\crcr\rightarrowfill\crcr
              \noalign{\kern-1pt\nointerlineskip}
              \hskip3.pt${\scriptstyle #1}$\hskip3.pt\crcr}}}
\def\otto{{\kern-1.truept\leftarrow\kern-5.truept\to\kern-1.truept}}
\def\arm{{}}
\font\bigfnt=cmbx10 scaled\magstep1

\newcommand{\card}[1]{\left|#1\right|}
\newcommand{\und}[1]{\underline{#1}}
\def\1{\rlap{\mbox{\small\rm 1}}\kern.15em 1}
\def\ind#1{\1_{\{#1\}}}
\def\bydef{:=}
\def\defby{=:}
\def\buildd#1#2{\mathrel{\mathop{\kern 0pt#1}\limits_{#2}}}
\def\card#1{\left|#1\right|}
\def\proof{\noindent{\bf Proof. }}
\def\qed{ \square}
\def\reff#1{(\ref{#1})}
\def\eee{{\rm e}}
\def\be{\begin{equation}}
\def\ee{\end{equation}}

\title{\Large A correction to a remark in a paper by Procacci and Yuhjtman:
new lower bounds for the  convergence radius of the virial series}

\author{\normalsize Aldo Procacci\footnote{\scriptsize Departamento de Matem{\'a}tica, Universidade Federal de Minas Gerais, Belo
Horizonte-MG, Brazil - aldo@mat.ufmg.br},
}

\maketitle

\begin{abstract}
In this note we
deduce a new lower bound for the convergence radius of the Virial series
of a continuous system of classical particles interacting via a stable and tempered pair potential
using the  estimates on the Mayer coefficients obtained in the  recent paper by Procacci and Yuhjtman (Lett Math Phys 107:31-46, 2017).
This corrects  the wrongly optimistic lower bound for the same radius
claimed (but not proved) in the above cited paper (in Remark 2 below Theorem 1).
The  lower bound for the convergence radius of the Virial series provided here represents a
strong improvement on the classical estimate given by Lebowitz and Penrose in 1964.
\vskip.3cm

%
%
\end{abstract}

\vskip.3cm
{\footnotesize
\\{\bf Keywords}: Classical continuous gas, Virial series.

\\{\bf MSC numbers}:  82B05, 82B21.
}

\numsec=1\numfor=1
\let\thefootnote\relax\footnotetext{2010 {\it Mathematics Subject Classification.} Primary 82B21; Secondary
05C05.}
\subsection*{1. Introduction}
\\The $n$-order coefficient of the Mayer series of a continuous system of  classical particles at inverse temperature $\b>0$ confined in a box
$\L\subset\mathbb{R}^d$ and interacting via a pair potential $V$ is explicitly given by
$$
C_n(\beta,{\Lambda} )~=~\cases{{1\over n!}{1\over |{\Lambda} |}\int\limits_{{\Lambda} }d{x}_1
\dots \int_{{\Lambda} } d{x}_n \sum\limits_{g\in G_{n}}~
\prod\limits_{\{i,j\}\in E_g}\left[  e^{ -\beta V({x}_i -{x}_j)} -1\right] & if $n\ge 2$\cr\cr
1 &if $n=1$
}\Eq(ursm)
$$
where $G_n$  is the set of the connected graphs in
$[n]$ and $E_g$ denotes the edge set of $g \in G_n$.
\\In our  recent paper  \cite{PY} it is
 proved that if the potential $V$ is stable and tempered
with stability constant $B$ (see (2.4) ahead)  the following upper bound holds.
$$
|C_n({\Lambda} ,\beta)|\le e^{\beta B n}{n^{n-2}\over n!} {[\tilde C(\beta)]^{n-1}}\Eq(Proyuh)
$$
where
$$
\tilde C(\beta)=\int_{\mathbb{R}^{d}} dx ~ \left(1-e^{-\beta |V(x)|}\right)
$$
This improves the classical estimates for the same coefficients obtained by Penrose and Ruelle in 1963 \cite{Pe63,Ru63}, namely
$$
|C_n({\Lambda} ,\beta)|\le e^{\beta 2B (n-2)}{n^{n-2}\over n!} {[C(\beta)]^{n-1}}\Eq(peru)
$$
with
$$
C(\beta)=\int_{\mathbb{R}^{d}} dx ~ \left|1-e^{-\beta V(x)}\right|
$$
\\The pressure $P_\L(\b,\l)$ and the density $\r_\L(\b,\l)$  of the system can be written  in terms of this coefficients as power series in the fugacity $\l$  as
$$
\b P_\L(\b,\l) ~= ~\sum_{n=1}^{\infty}C_n(\b,\L)\l^n\Eq(pressu)
$$
$$
\r_\L(\b,\l)~=~\l{\partial\over \partial\l}(\b P_\L(\b,\l))~=~\sum_{n=1}^{\infty}nC_n(\b,\L)\l^n\Eq(densi)
$$
and by estimates \equ(Proyuh) one immediately  concludes that $P_\L(\b,\l)$ and  $\r_\L(\b,\l)$ are analytic functions of  $\l$ for all complex values of $\l$ satisfying
$$
|\l|< R^*\doteq {1\over e^{\b B+1}\tilde C(\b)}
$$

\\The  lower bound $ R^*$ for the convergence radius of the Mayer series obtained in
\cite{PY}
improves the one given by Ruelle and Penrose in 1963 \cite{Pe63,Ru63} derived from estimates \equ(peru).

\\By eliminating the activity $\l$ from equation \equ(pressu) and \equ(densi) one can write the pressure of the system in the grand canonical ensemble in power of the
density $\r=\r_\L(\b,\l)$ obtaining the so-called { Virial expansion of the pressure}, which   is usually written as
$$
\b P_\L(\b,\r)= \r - \sum_{k\ge 1} {k\over k+1} \b_k(\b,\L)\r^{k+1}\Eq(virial)
$$
The coefficients $ \b_k(\b,\L)$  are  of course   certain algebraic combinations  of the Mayer coefficients $C_n(\b,\l)$ with
$n\le k+1$ whose explicit expression
is long known (see e.g.   formula (29) p. 319 of \cite{PB} and references therein). In 1964  Lebowitz and Penrose developed an indirect method, based on Lagrange inversion, to derive a lower bound for
the convergence radius $\RR$ of the Virial series \equ(virial) from  the  Penrose-Ruelle estimates \equ(peru). Namely, they proved that the r.h.s. of \equ(virial) converges absolutely for all complex $\r$ such that
$$
|\r|<  {\cal R}_{\rm LP}\doteq {g(e^{2\b B})\over e^{2\b B} C(\b)}\Eq(virialold)
$$
with the function
$$
g(u)= \max_{0<w<\ln(1+u)} {[(1+u)e^{-w} -1]w\over u}\Eq(gi)
$$
slightly increasing in the interval $[1,+\infty)$ and  such that $g(1)\approx 0.14477$ and $\lim_{u\to\infty }g(u)= e^{-1}$.

\\Our paper \cite{PY} contains a remark (Remark 2,
immediately after Theorem 1)
claiming that
the new bounds for the Mayer coefficients \equ(Proyuh) also yields an improvement (formula (2.17) in \cite{PY}) on   $\RR_{\rm LP}$ . In \cite{PY} it is alleged that to get
this new lower bound of the
convergence radius of the virial series one has just to
redo  the calculations performed by Lebowitz and Penrose in \cite{LP}  using the
new upper bound  of the n$^{\rm th}$-order Mayer coefficients \equ(Proyuh) in place  of the old ones  \equ(peru) given by Penrose and Ruelle.
This assertion is wrong since the new bounds of the $n$-order Mayer coefficients \equ(Proyuh) actually improve on \equ(peru) only as soon as  $n\ge 4$, and it happens that
the $n=2$ and the $n=3$ order  Mayer coefficients have a non-negligible influence in the deduction
presented in \cite{LP}.
The purpose of this note is thus to correct this error and explain in details how to deduce
a  new lower bound of the convergence radius
of the virial series (which still strongly improves on \equ(virialold))
from  (a slight variant of) the new upper bounds of the Mayer coefficients  \equ(Proyuh) and via  the method used in \cite{LP}.


\vskip.51cm

\subsection*{2. The new lower bound for the convergence radius of the Virial series}
\numsec=2\numfor=1
\\Let us first of all introduce some notations and definitions. Given a pair potential $V$ we recall
that
$$
 B=\sup_{n\ge 2}~~\sup_{(x_1,\dots,x_n)\in \mathbb{R}^{dn}}\Big\{-{1\over n}\sum_{1\le i<j\le n}V(x_i-x_j)\Big\}\Eq(stabs)
$$
is the usual stability constant of $V$.
We also set
$$
\bar B=\sup_{n\ge 2}~~\sup_{(x_1,\dots,x_n)\in \mathbb{R}^{dn}}\Big\{-{1\over n-1}\sum_{1\le i<j\le n}V(x_i-x_j)\Big\}\Eq(bSatbs)
$$
and call $\bar B$ the {\it Basuev stability constant} of the potential $V$  (after Basuev who was the first to introduce it in \cite{Ba1}).
We have clearly, for all $n\ge 2$ and all $(x_1,\dots,x_n)\in \mathbb{R}^{dn}$
$$
\sum_{1\le  i<j\le n} V(x_i-x_j)\ge -n B \Eq(stab)
$$
$$
\sum_{1\le i<j\le n} V(x_i-x_j)\ge -(n-1) \bar B \Eq(stabas)
$$
and
$$
\bar B\ge B
$$
For the majority of realistic stable potentials the constants $\bar B$ and B are likely to be  very close (if not equal). E.g.
for the Leonard-Jones potential in three dimensions $V_{\rm LJ}(x)={|x|^{-12}}-{2 |x|^{-6}}$, according to the tables given in \cite{JI}, we have that $B\le \bar B\le {1001\over 1000}B$. In any case,
that for general stable potentials  it holds that
$$
\bar B\le {d+1\over d}B
\Eq(basgen)
$$
while for potentials which
reach
a  negative minimum at some  $|x|=r_0$
and are negative for all $|x|>r_0$ (e.g. Lennard-Jones type potentials) it holds that
$$
\bar B\le \cases{{3\over 2} B & if $d=1$\cr \cr
{7\over 6} B & if $d=2$\cr \cr
{2d(d-1)+1\over 2d(d-1)} B & if $d\ge 3$}\Eq(estbar)
$$

\\Now, using \equ(stabas) in place of \equ(stab), with minor changes in the proof given in \cite{PY} one can show that the inequality  \equ(Proyuh) can be rewritten in terms of the Basuev constant $\bar B$ as follows
$$
|C_n({\Lambda} ,\beta)|\le e^{\beta \bar B (n-1)}n^{n-2} {[\tilde C(\beta)]^{n-1}\over n!}\Eq(Proyuh2)
$$
Using these bounds \equ(Proyuh2) and  following the
strategy described in \cite{LP} we have
the following Theorem.
\begin{teorema}\label{pryu}
 Let $V$ be a stable and tempered pair  potential with  Basuev stability constant $\bar B$. Then
the convergence radius $\RR$ of the Virial  series \equ(virial)  admits the following lower bound.
$$
\RR\ge \RR^*\doteq  { g(1)\over  \tilde C(\b) e^{\b \bar B}}\Eq(PYvi)
$$
where $g$ is the function defined in \equ(gi).
\end{teorema}
Note that the difference with the incorrect announced bound of formula (2.17) of  \cite{PY}  is that in the correct expression \equ(PYvi)
the Basuev stability constant $\bar B$ replaces the usual stability constant $\bar B$ and moreover \equ(PYvi) contains a $g(1)$, rather than a $g(e^{ \beta \bar B})$.
We will prove Theorem \ref{pryu} in the next section and,
to make this note as self-contained as possible, we will also sketch,  in  Appendix $\rm A$, the proof of \equ(Proyuh2) while in Appendix $\rm B$ we will prove bounds \equ(basgen) and \equ(estbar).  \

\\We conclude this section by comparing    the new bound $\RR^*$ \equ(PYvi) with the Lebowitz Penrose bound $\RR_{\rm LP}$ \equ(virialold).
Let us first observe that the ratio ${\RR^*/ \RR_{\rm LP}}$  is 1 for positive potentials, so there is no improvement when  $B=\bar B=0$.
On the other hand  for potentials with stability constant $B>0$ (i.e. for potentials
with a negative part) the ratio $\RR^*/\RR_{\rm LP}$ grows exponentially in $\b$, since so do ${C(\b)/\tilde C(\b)}$ and
$e^{\b(2B-\bar B)}$.
 To give an idea of how significant  can be the  improvement  in a concrete example, let us look at
the case of the three-dimensional Lennerd-Jones potential $V_{\rm LJ}(x)={|x|^{-12}}-{2 |x|^{-6}}$ considered in
\cite{PY}. We have that
at inverse temperature $\b=1$, using the value $B_{\rm LJ}=8.61$
for its stability constant  and the fact that $\bar B_{\rm LJ}\le {1001\over 1000}B_{\rm LJ}$   (see \cite{JI}),  $\RR^*$
is at least $3.3\times 10^4$ larger than $\RR_{\rm LP}$, while for $\b=10$
$\RR^*$ is at least $2.9\times 10^{43}$ larger than $\RR_{\rm LP}$.

\subsection*{3. Proof of Theorem 1}\label{subsec3}
\\We start by observing that, due to \equ(densi) and the fact that $C_1(\b,\L)=1$, there exists a circle $C$
 of some radius $R< 1/[\tilde C(\b)e^{\b B+1}]$ and center in the origin  $\l=0$ of the complex $\l$-plane such that
  $\r_\L(\b,\l)$
has only one zero  in the disc $\bar D_R=\{\l\in \C: |\l|\le R\}$ and this zero occurs precisely at $\l=0$.
  Let now  $\r\in \C$ be such that
$$
|\r|< \min_{\l\in C}|\r_\L(\b,\l)| \Eq(condm)
$$
Then by Rouch\'e's theorem $\r_\L(\b,\l)$ and
 $\r_\L(\b,\l)-\r$ have the same number of zeros (i.e. one)  in the region $D_R=\{\l\in \C: |\l|\le R\}$. In other words, for any complex $\r$ satisfying
  \equ(condm) there is only one $\l\in D_R$ such that  $\r=\r_\L(\b,\l)$ and therefore we can invert the equation   $\r=\r_\L(\b,\l)$ and write
 $\l=\l_\L(\b, \r)$. Thus, according to  Cauchy's argument principle, we can write the pressure $\b P_\L(\b,\l)$
  as a function of the density $\r=\r_\L(\b,\l)$  as
$$
P_\L(\r,\b)={1\over 2\p i}\oint_\g P_\L(\b,\l){d\r_\L(\b,\l)\over d\l}{d\l\over \r_\L(\b,\l)-\r}\Eq(cauchy)
$$
where $\g$ can be  any circle centered at the origin in the  complex $\l$-plane fully contained in the region $D_R$
and such that
$$
|\r|<\min_{\l\in \g} | \r_\L(\b,\l)|\Eq(minro)
$$
By standard complex analysis  $P_\L(\r,\b)$ is analytic in $\r$ in the region \equ(minro). Indeed,
once \equ(minro) is satisfied we can write
$$
{1\over \r_\L(\b,\l)-\r}=\sum_{n=0}^{\infty} {\r^n\over  [\r_\L(\b,\l)]^{n+1}}\Eq(okso)
$$
and inserting \equ(okso) in \equ(cauchy) we get
$$
P_\L(\r,\b)= \sum_{n=1}^\infty c_n(\b,\L)\r^n\Eq(virial3)
$$
with
$$
c_n(\b,\L)= {1\over 2\p i}\oint_\g P_\L(\b,\l){d\r_\L(\b,\l)\over d\l}{d\l\over  [\r_\L(\b,\l)]^{n+1}} ~=
~
~-{1\over 2\p in}\oint_\g P_\L(\b,\l){d\over d\l}\left[{1\over  [\r_\L(\b,\l)]^{n}}\right]d\l~=
$$
$$
= {1\over 2\p in}\oint_\g {dP_\L(\b,\l)\over d\l}{1\over  [\r_\L(\b,\l)]^{n}}d\l~= ~ {1\over 2\p in\b} \oint_\g{1\over
[\r_\L(\b,\l)]^{n-1}}{d\l\over \l}~~~~~~~~~~~~~~~~~~~~~
$$
Therefore
$$
|c_n(\b,\L)|\le{1\over n\b}{1\over \left[\min_{\l\in \g} |\r_\L(\b,\l)|\right]^{n-1}}\Eq(cn)
$$
Inequality \equ(cn) shows that the convergence radius $\RR$ of the series \equ(virial3) (i.e. of the virial series \equ(virial)) is such
that
$$
\RR\ge \min_{\l\in \g}   |\r_\L(\b,\l)|\Eq(Ok)
$$
Therefore  the game is to find an optimal circle $\g$ in the region $D_R$ which maximizes  the r.h.s. of \equ(Ok).
We proceed as follows.
Recalling \equ(densi)  we have, by the triangular inequality, that
$$
 |\r_\L(\b,\l)| \ge |\l|- \sum_{n=2}^\infty n|C_n(\b,\L)||\l|^n \Eq(triangul)
$$
We now  use estimate \equ(Proyuh2) to bound the sum in the r.h.s. of  \equ(triangul).
Therefore we get
$$
 |\r_\L(\b,\l)| ~\ge ~|\l|- \sum_{n=2}^\infty {n^{n-1}\over n!}[\tilde C(\b) e^{\b \bar B}]^{n-1}|\l|^n~=~
 2|\l|-{1\over \tilde C(\b) e^{\b \bar B}}\sum_{n=1}^\infty {n^{n-1}\over n!}[\tilde C(\b) e^{\b \bar B}|\l|]^{n}
$$
Now following \cite{LP} let us set $w$ to be the  first positive solution of
$$
we^{-w}= \tilde C(\b) e^{\b \bar B}|\l|\Eq(wew)
$$
If we  take $|\l|< 1/ e^{\b \bar B+1} \tilde C(\b)$ (which is
surely  inside the convergence region since $\bar B\ge B$) then $\tilde C(\b) e^{\b \bar B}|\l|<1/e$ and since the function $we^{-w}$ is
increasing in
the interval $[0,1]$ and takes the value $1/e$ at $w=1$,  there is a unique $w$ in the interval $[0,1]$ which solves \equ(wew).
Now we use the Euler's formula
$$
w=\sum_{n=1}^\infty{n^{n-1}\over n!}(we^{-w})^n
$$
to get
$$
|\r_\L(\b,\l)|\ge {w\over \tilde C(\b) e^{\b \bar B}}\left[2e^{-w}
-1\right]\Eq(fififi)
$$
Observe that the r.h.s. \equ(fififi) is greater than zero when $w$ varies in the interval  $(0,\ln 2)$. Therefore, any  circle $\g$ around the origin with radius $R_\g$ between zero and
$\ln 2/(2e^{\b \bar B+1}\tilde
C(\b))$,  which  corresponds, in force of \equ(wew),  to  any  $w$  in the interval $(0,\ln 2)$, is surely in the region $D_R$. So
 we have
$$
\max_{\g\subset D_R} \min_{\l\in \g}   |\r_\L(\b,\l)| \ge \max_{w\in (0,\ln2)} {w\over \tilde C(\b) e^{\b \bar B}}\left[2e^{-w}
-1\right]\Eq(fine)
$$
which, recalling  \equ(Ok) and \equ(gi), concludes  the proof.

\vskip.55cm
\renewcommand{\theequation}{A.\arabic{equation}}
\setcounter{equation}{0}  
\subsection*{Appendix A. Proof of \equ(Proyuh2)}
Estimate \equ(Proyuh2) is basically the bound \equ(Proyuh) proved in \cite{PY} with the unique difference that the factor $e^{\b \bar B(n-1)}$ replaces
the factor  $e^{\b  B n}$. Of course, since $\bar B\ge B$, this is a bad deal for  someone interested in an upper bound  the Mayer series. On the other hand, as shown in Sec. 3, the use \equ(Proyuh2) instead of  \equ(Proyuh) happens to be a  good deal towards an upper bound  the Virial series.
Inequality  \equ(Proyuh2) relies on two lemmas originally proved in \cite{PY} (see there Lemma 1 and Lemma 2). For completeness we report
these lemmas and their proofs here below.

\\The first of these two lemmas involves the concept of partition scheme, which, we remind, is a map
$\bm{M}$ from the set $T_n$  of the labeled trees  in   $[n]$ to the set $G_n$ of the connected graphs  in $[n]$  such that
$G_n=\biguplus_{\tau\in {T_n}}[\tau,{\bm M}(\tau)]$
with $\biguplus$  disjoint union and $[\tau,\bm M(\tau)]=\{g\in G_n: \tau\subseteq g\subseteq \bm M(\tau)\}$.

\begin{lema}\label{parti} For fixed  $V$ and $(x_1,\dots, x_n)\in \mathbb{R}^{dn}$,  choose a total order $\succ$ in the set of  edges $E_n$ of the complete graph $K_n$
in such a way that $\{i,j\}\succ\{k,l\}\Longrightarrow  V(x_i-x_j)\ge V(x_k-x_l)$
 and let $\bm T:G_n\to T_n$   be the map  that
associates to $g\in G_n$ the tree  $\bm T(g)\in T_n$ constructed by  starting from $\emptyset$ and keeping adding
the lowest edge in $g$  that does not create a cycle (Kruskal algorithm).

\\Let $\bm M: T_n\to G_n$ be the map that associates to  $\tau\in T_n$ the graph $\bm M( \tau )\in G_n$
whose edges are the $\{i,j\}\in E_n$
such that $\{i,j\} \succeq \{k,l\} $ for every edge $\{k,l\} \in E_\tau$ belonging to the path  from $i$ to $j$ through $\tau$.

 \\Then  $\bm T^{-1}(\tau)=\{g \in {\mathcal G}_n :\, \tau \subseteq g \subseteq \bm M(\tau)\}$
 and therefore $\bm M$  is a partition scheme in ${G}_n$.
\end{lema}
{\bf Proof}.
Assume first that $g \in \bm T^{-1}(\tau)$. Then $\tau=\bm T(g) \subset g$. Now take $\{i,j\} \in E_g$, and let $e \in E_\tau$
 be any edge belonging to the path from $i$ to $j$ in $\tau$.
 Consider the tree $\tau'$  obtained from $\tau$ after replacing the edge $e$ by $\{i,j\}$.
 By minimality of $\tau$ we must have $ \{i,j\}\succ e$, i.e. $\{i,j\} \in E_{\bm M(\tau)}$, whence $g \subset \bm M(\tau)$.
Conversely, let $\tau \subset g \subset \bm M(\tau)$. We must show $\bm T(g)=\tau$.
By contradiction, take $\{i,j\} \in E_{\bm T(g)} \setminus E_\tau$. Consider the path $p^\tau(\{i,j\})$ in $\tau$ joining $i$ with $j$.
   Since $\bm T(g) \subset \bm M(\tau)$, $\{i,j\}$ is greater (w.r.t. $\succ$) than  any
 edge in the path $p^\tau(\{i,j\})$. If we remove $\{i,j\}$ from $\bm T(g)$, the tree $\bm T(g)$ splits into two trees.
 Necessarily,  one of the edges in
 the path $p^\tau(\{i,j\})$ joins a vertex of one tree with a vertex of the other. Thus, by adding this edge we get   a new tree
 which contradicts the minimality of $\bm T(g)$. $\Box$
\vskip.1cm
\\{\bf Remark}. The proof of Lemma 1 above, as well as the definition of the partition scheme
$\bm M$, are identical to those given in the homonymous lemma of \cite{PY}, but the  map $\bm T: G_n\to T_n$
appearing in the enunciate, based on Kruskal's algorithm,  replaces a similar (but more involved) map constructed in \cite{PY} via so-called admissible functions with values in a tomonoid. The idea to use the Kruskal's algorithm, which eases
the definition of the ``minimal tree" map $\bm T$, has been suggested  during an Oberwolfach meeting by David Brydges, Tyler Helmuth and Daniel Ueltschi (see \cite{Ue}).

\noindent
Using the partition scheme $\bm M$ defined in Lemma \ref{parti} we now state and prove the second  key lemma of \cite{PY}.
\begin{lema}\label{stabp}
 Let $V$ be  stable  with Basuev stability constant $\bar B$, and let $\tau \in { T}_n$.
Let  $(x_1,...,x_n) \in {\mathbb R}^{dn}$ and let $\bm M$ be the partition scheme given above, then
\begin{equation}\label{Peb}
 \sum_{\{i,j\} \in \bm E_{M(\tau)} \setminus E_\tau^+} v(x_i-x_j) \geq -\bar B(n-1)
\end{equation}
\end{lema}
{\bf Proof}.
 The set of edges $E_\tau\setminus E_\tau^+$ forms  the forest $\{\tau_1,...,\tau_k\}$.
 Let us denote ${ V}_{\tau_s}$ the vertex set of the  tree $\tau_s$
 from the forest.
Assume $i \in { V}_{\tau_a}$, $j \in {V}_{\tau_b}$.
 If $a \neq b$, the path from $i$ to $j$ through $\tau$ involves an edge $\{k,l\}$ in $E_\tau^+$. Thus, if in addition $\{i.j\} \in
E_{\bm M(\tau)}$, we have $\{i,j\}\succeq \{k,l\}$ and therefore
 $v(x_i-x_j) \geq v(x_k-x_l) \geq 0$. If $a=b$, the path from $i$ to $j$ through $\tau$ is contained in $\tau_a$. Thus, if in addition
 $\{i,j\} \notin E_{\bm M(\tau)}$, there must be at least  one edge $\{r,s\}$ in that path such that
 $\{i,j\}\prec \{r,s\}$  and  therefore  $v(x_i-x_j) \leq v(x_r-x_s) < 0$.  This allows to bound:

 $$\sum_{\{i,j\} \in E_{\bm M(\tau)} \setminus E_\tau^+}v(x_i-x_j) \geq \sum_{s=1}^k \sum_{\{i,j\} \subset { V}_{\tau_s}} v(x_i-x_j) \geq
 \sum_{s=1}^k -| (V_{\tau_s}|-1)\bar B \ge -(n-1)\bar B$$
where to get the last inequality we have used \equ(stabas). $\Box$

\vskip.2cm

\\We are now ready to conclude the proof of \equ(Proyuh2).
By the  so-called Penrose tree-graph identity \cite{Pe67}, given a pair potential $V$ and  $(x_1,...,x_n) \in {\mathbb R}^{dn}$, one has that
\begin{equation}\label{r.200}
\sum_{g\in G_{n}}~
\prod_{\{i,j\}\in E_g}\left[  e^{ -\beta V(x_i-x_j)} -1\right]=
\sum_{\tau\in T_n}
e^{-\beta\sum_{\{i,j\}\in E_{\bm{M}(\tau)}\backslash E_\tau}V(x_i-x_j)}
\prod_{\{i,j\}\in E_\tau}\left(e^{- \beta V(x_i-x_j)}-1\right)
\end{equation}
where $\bm{M}: T_n\to G_n$  is any map  such that
$G_n=\biguplus_{\tau\in {T_n}}[\tau,{\bm M}(\tau)]$
with $\biguplus$  disjoint union and $[\tau,\bm M(\tau)]=\{g\in G_n: \tau\subseteq g\subseteq \bm M(\tau)\}$  ({\it partition scheme})
and ${ T}_n$ is the set of all trees with vertex set $[n]$.
\vskip.12cm
\noindent
Set now $E_\tau^+=\{\{i,j\}\in E_\tau:V(x_i-x_j)\ge 0\}_{}$, then
from (\ref{r.200})
one immediately gets
\begin{equation}\label{r.200b}
\left|\sum\limits_{g\in G_{n}}~
\prod\limits_{\{i,j\}\in E_g}\left[  e^{ -\beta V(x_i-x_j)} -1\right]\right|~\le
~
\sum_{\tau\in T_n}
e^{-\beta\sum\limits_{\{i,j\}\in E_{\bm{M}(\tau)}\backslash E^+_\tau}V(x_i-x_j)}
\prod_{\{i,j\}\in E_\tau}\left(1-e^{- \beta |V(x_i-x_j)|}\right)
\end{equation}

\vskip.2cm
\noindent
Let us now choose the partition scheme defined in Lemma \ref{parti}. Then
from Lemma \ref{stabp}, inserting (\ref{Peb})
into (\ref{r.200b}), one obtains, for any
$n\ge 2$ and any $(x_1,\dots,x_n)\in \mathbb{R}^{dn}$,  the following inequality
\begin{equation}\label{from}
|\sum_{g \in {G}_n} \prod_{\{i,j\} \in E_g} (e^{-\beta V(x_i-x_j)}-1)|~\le ~
e^{\beta \bar B (n-1)}\sum_{\tau \in {T}_n} \prod_{\{i,j\} \in E_\tau} (1-e^{-\beta| V(x_i-x_j)|})
\end{equation}

\\Now  \equ(Proyuh2)  follows easily from (\ref{from}) recalling  that $|T_n|=n^{n-2}$ and observing that, for any  $\tau\in T_n$,  it holds
\begin{equation}\label{treintg}
\int_{{\Lambda} }d{x}_1\dots \int_{{\Lambda} }d{x}_{n} \prod_{\{i,j\}\in E_{\tau}}
\left(1-e^{-\beta |v(x_i-x_j)|}\right)\le
|{\Lambda} |\left[\tilde C(\beta)\right]^{n-1}
\end{equation}

\vskip.55cm
\renewcommand{\theequation}{B.\arabic{equation}}
\setcounter{equation}{0}  
\subsection*{Appendix B. Proof of \equ(basgen) and \equ(estbar)}

\\{\bf Proof of \equ(basgen)}.
Let  assume that $V(|x|)$ is  a stable and tempered pair potential  in $d$ dimensions with stability constant $B$ and Basuev stability constant $\bar B$.
Let us first prove that if
\be\label{supbar}
\bar B= \limsup_{n\to \infty} \bar B_n
\ee
then
$$
\bar B = B
$$
Suppose by contradiction that $\bar B- B=\d>0$. If this holds
then necessarily there exists a finite  $m \ge 2$ such that $B=B_m$ otherwise if $B=\limsup_{n\to \infty} B_n$ then $B=\bar B$.
Now, due to the hypothesis (\ref{supbar}), for all $\e>0$ there exists $n_0$ such that
for infinitely many $n>n_0$
$$
\bar B_n> \bar B-\e~~~~~~~~~~\Longrightarrow ~~~~~~ B_n> {n-1\over n}(\bar B-\e)= {n-1\over n}(B+\d-\e)~
$$
Choose  $\e={\d\over 2}$, then  for infinitely many $n$ we have that
$$
B_n>  {n-1\over n}(B+{\d\over 2}) > B~~~~~~~~~~~~\mbox{ as soon as $n> {2B\over \d}+1$}
$$
in contradiction with the assumption that  $B=B_m$. Hence we must have $B=\limsup B_n= \limsup \bar B_n=\bar B$.

\\So let us suppose that the $\sup \bar B_n$ is reached at some finite integer $m$, i.e. $\bar B=
\bar B_m$.

\\Since $V$ is  stable, it is bounded from below and since $V$ is tempered $\inf V$ cannot be positive.
Let $\inf_{x\in \mathbb{R}^d}V(x)=-C$ with $C\ge 0$. Then for any $\e>0$ there exists $r_\e$ such that
$V(r_\e)< -(C -\e)$.
Take the configuration $(x_1,x_2,\dots, x_{d+1})\in (\mathbb{R}^d)^{d+1}$ such that $x_1$, $x_2,\dots ,x_{d+1}$ are vertices of a
$d$-dimensional
hypertetrahedron  with sides of length $r_\e$. Recall that a $d$-dimensional
hypertetrahedron has $d+1$ vertices and $d(d+1)/2$ sides.
Then $U(x_1,x_2,\dots, x_{d+1})< -{d(d+1)\over 2}(C-\e)$, which implies that $B> {d\over 2}(C-\e)$ and by the arbitrariness of $\e$ we get
$B\ge {d\over 2}C$.
On the other hand we also have, for any $(x_1,\dots,x_m)\in \mathbb{R}^{dm}$ that
$U(x_1,\dots,x_m)\ge -{m(m-1)C/2}$ which implies that $\bar B_m=\bar B\le {m\over 2}C$. Hence  we can write ${d\over 2}C\le B\le \bar
B=\bar B_m\le {m\over 2}C$ which implies  $m\ge d+1$ and so   $\bar B= \bar B_m= {m\over m-1}B_m\le {m\over m-1}B\le {d+1\over d}B$. $\Box$

\vskip.3cm
\\{\bf Proof of \equ(estbar)}.
Let  us  assume that $V(|x|)$ is  a stable pair potential in $d$ dimensions with stability constant $B$ and Basuev stability constant $\bar B$ and that  $V(|x|)$ reaches
its  negative minimum $-C$ at some  $|x|=r_0$
and it is negative for all $|x|>r_0$. We want to prove that the inequalities  \equ(estbar) hold.
\\

\\First note that $d(d-1)C$ is always a lower bound for $B$ when $d\ge 3$. Just consider a configuration in which $n$ particle
 (with $n$ as large as we want) are
arranged in close-packed configuration  at the sites of a $d$-dimensional  face-centered cubic lattice with step $r_0$. The energy of such configuration
is (asymptotically
as $n\to \infty$)
less than or equal to $-d(d-1)Cn$ since
in a $d$-dimensional face-centered cubic lattice each site has $2d(d-1)$ neighbors (see e.g. \cite{TJ}) and so there are  (asymptotically)
$d(d-1)n$ pairs of neighbors in the configuration.
On the other hand,
for any $n$-particle configuration $(x_1,\dots,x_n)\in \mathbb{R}^{dn}$,
it holds that $U(x_1,\dots,x_n)\ge -n(n-1)C/2$, i.e. $\bar B_n\le nC/2$. Now,
 if $\bar B= \sup_n\bar B_n$ is attained at $n\to \infty$ then, as previously seen, we have $\bar B=B$. So let us suppose
 that $\bar B= \sup_n\bar B_n$ is attained at a certain finite $m$. Then
we must have that  $d(d-1)C\le B\le \bar B\le mC/2$ and so $m \ge 2d(d-1)$. Now if $m=2d(d-1)$ then $\bar B=B$. So when $\bar B>B$ then
$m > 2d(d-1)$ and so we have $\bar B=\frac{m}{m-1}B_m \leq {2d(d-1)+1\over 2d(d-1)} B$. The case $d=1$ and $d=2$ are treated analogously by just
observing that the close-packed arrangement in $d=1$ is simply the cubic lattice with $2$ neighbors for each site while for $d=2$ is the
triangular lattice with $6$ neighbors for each site. So $d(d-1)$ must be replaced by $1$ for $d=1$ and by $3$ for $d=2$ yielding $m>2$ and
$m>6$ for $d=1$ and $d=2$ respectively. $\Box$

\vskip.51cm

\subsection*{Acknowledgment}
It is a pleasure to thank Sergio Yuhjtman for his reading of the manuscript and his useful observations and suggestions.
This work has been partially supported by the Brazilian  agency CNPq
(Conselho Nacional de Desenvolvimento Cient\'{\i}fico e Tecnol\'ogico - Bolsa de Produtividade em pesquisa, grant n. 306208/2014-8).

\vskip.55cm

\renewcommand{\section}[2]{}%


\subsection*{References}
\begin{thebibliography}{99}











\bibitem{Ba1}  A. G. Basuev (1978) :  {\it A theorem on minimal specific energy for classical systems}. Teoret. Mat. Fiz.
{\bf 37}, no. 1, 130--134.

%








%






%

\bibitem{LP}  J. L. Lebowitz and O. Penrose (1964): {\it Convergence of Virial Expansions}, J. Math. Phys. {\bf 7}, 841-847.



%



\bibitem{JI} J. E. Jones; A. E. Ingham (1925): {\it On the calculation of certain crystal potential constants, and on the cubic
crystal of least potential energy}. Proc. Roy. Soc. Lond. A {\bf 107}, 636--653.




%

%










%





\bibitem{PB} R. K. Pathria and P. D. Beale (2011): {\it Statistical mechanics, Third edition}, Elsevier, Amsterdam.

\bibitem{Pe63} O. Penrose (1963): {\it Convergence of Fugacity Expansions for Fluids and Lattice Gases}, J. Math. Phys. {\bf 4},  1312 (9
    pages).


\bibitem{Pe67} O. Penrose (1967): {\it Convergence of fugacity
 expansions for classical systems}.  In {\it Statistical
 mechanics: foundations and applications}\/, A. Bak (ed.), Benjamin, New York.






\bibitem{PY}   A. Procacci and S. A. Yuhjtman (2017): {\it  Convergence of Mayer and Virial expansions
and the Penrose tree-graph identity}, Lett. Math. Phys., {\bf 107}, 31--46 (2017).








\bibitem{Ru63} D. Ruelle (1963): {\it Correlation functions of classical gases}, Ann. Phys., {\bf 5}, 109--120.






\bibitem{TJ} S. Torquato; Y. Jiao: {\it  Effect of dimensionality on the percolation thresholds of various d-dimensional lattices}
Phys.  Rev.  E {\bf 87}, 032149 (2013).


\bibitem{Ue} D. Ueltschi (2017): {\it An improved tree-graph bound}. To appear in Oberwolfach Reports, arXiv:1705.05353.





\end{thebibliography}
\end{document}